\documentclass[11pt,,a4paper,onecolumn,draftcls]{IEEEtran}
\usepackage{graphicx}
\usepackage{epsfig}
\usepackage{subfigure}
\usepackage{amssymb}
\usepackage{amsbsy}
\usepackage{amsmath}
\usepackage{latexsym}       % to get LASY symbols
\usepackage{epsf}
\newcommand{\ve}{\mbox{${\mathbf e}$}}

\newcommand{\vs}{\mbox{${\mathbf s}$}}

\newcommand{\vx}{\mbox{${\mathbf x}$}}

\newcommand{\vy}{\mbox{${\mathbf y}$}}

\newcounter{theorem}
\newtheorem{theorem}{Theorem}
\newtheorem{lemma}[theorem]{Lemma}
\newtheorem{claim}[theorem]{Claim}
\newcounter{definition}
\newtheorem{definition}{Definition}

\newcommand{\beq}{\begin{equation}}
\newcommand{\eeq}{\end{equation}}
\newcommand{\bea}{\begin{array}}
\newcommand{\ena}{\end{array}}
\newcommand{\bds}{\begin {itemize}}
\newcommand{\eds}{\end {itemize}}
\newcommand{\bdf}{\begin{definition}}
\newcommand{\blm}{\begin{lemma}}
\newcommand{\edf}{\end{definition}}
\newcommand{\elm}{\end{lemma}}
\newcommand{\bthm}{\begin{theorem}}
\newcommand{\ethm}{\end{theorem}}
\newcommand{\bprp}{\begin{prop}}
\newcommand{\eprp}{\end{prop}}
\newcommand{\bcl}{\begin{claim}}
\newcommand{\ecl}{\end{claim}}
\newcommand{\bcr}{\begin{coro}}
\newcommand{\ecr}{\end{coro}}
\newcommand{\bquest}{\begin{question}}
\newcommand{\equest}{\end{question}}

\newcommand{\larrow}{{\larrow}}

\begin{document}
\title{Turbo Analog Error Correcting Codes Decodable By Linear Programming}
\author{Avi~Zanko$^{1,*}$,
			  Amir~Leshem$^{1}$,~\IEEEmembership{Senior Member,~IEEE,}
			  Ephraim~Zehavi$^{1}$~\IEEEmembership{Senior Member,~IEEE,}
\thanks{$^1$ School of Engineering, Bar-Ilan University, Ramat-Gan, 52900,Israel}
\thanks{$^*$ Corresponding author, email: AviZanz@gmail.com}
}
\date{}

\maketitle

%-------------abstract----------------------------------
%-------------------------------------------------------
\begin{abstract}
\label{sec:abstract}
In this paper we present a new Turbo analog error correcting coding scheme for real valued signals that are corrupted by impulsive noise. This Turbo code improves Donoho's deterministic construction by using a probabilistic approach. More specifically, our construction corrects more errors than the matrices of Donoho by allowing a vanishingly small probability of error (with the increase in block size). The problem of decoding the long block code is decoupled into two sets of parallel Linear Programming problems. This leads to a significant reduction in decoding complexity as compared to one-step Linear Programming decoding.

%In this paper we present a new error correction approach for signals with real valued entries. The goal is to recover an input codeword from its corrupted measurements. An analog Turbo Block Code based on a Given analog code is constructed to correct errors caused by impulsive noise using Linear Programming iteratively. This provide a deterministic construction improving to Donoho construction at the price of allowing arbitrary low probability of error that decaying sub exponentially to zero. Moreover, this construction leads to a much lower complexity compared to the direct approach.

%In this paper we discuss error correcting problem with real valued entries. We wish to recover an input vector $m \in R^k$ from the corrupted measurements $y = Gm+e$ .Where $\bf{G}$ is $n$ by $k$ code generator matrix and $e \in R^n$ is an arbitrary (sparse) error vector. Assume that we are given a code that is able to correct $\alpha n^{1/2}$ errors i.e. whenever $\left\|e\right\|_{\ell_0}\leq\alpha n^{1/2}$ we are able to correctly compute m from $y=Gm+e$. We show that it is possible to improve the code so that it corrects up to $\frac{\alpha n^{3/4}}{log n}$ errors with probability $1-\epsilon(n)$ where $\epsilon(n)$ decaying sub exponentially to zero. This provide a deterministic construction improving to Donoho construction at the price of allowing arbitrariy low probability of error.

\end{abstract}
%-------------end abstract------------------------------
\begin{IEEEkeywords}
Analog codes, Compressed Sensing, Linear Programming, Turbo decoding.
\end{IEEEkeywords}
%Analog codes, decoding of linear codes, Linear Programming, $\ell_1$ minimization, restricted isometry property, UUP, sparse solution to underdetermined systems, Compressed Sensing, Turbo decoding, product codes.
%%%%%%%%%%%%%%%%%%%%%%%%%%%%%%%%%%%%%%%%%%%%%%%%%%%%%%%%%%%%%%%%%%%%%%%%%%%%%%%%%%%%%%%%%%%%%%%%%%%%%%%%%%%%%%
%%%%%%%%%%%%%%%%%%%%%%%%%%%%%%%%%%%%%%%%%%%%%%%%%%%%%%%%%%%%%%%%%%%%%%%%%%%%%%%%%%%%%%%%%%%%%%%%%%%%%%%%%%%%%%
%%%%%%%%%%%%%%%%%%%%%%%%%%%%%%%%%%%%%%%%%%%%%%%%%%%%%%%%%%%%%%%%%%%%%%%%%%%%%%%%%%%%%%%%%%%%%%%%%%%%%%%%%%%%%%
%%%%%%%%%%%%%%%%%%%%%%%%%%%%%%%%%%%%%%%%%%%%%%%%%%%%%%%%%%%%%%%%%%%%%%%%%%%%%%%%%%%%%%%%%%%%%%%%%%%%%%%%%%%%%%
\section{Introduction}
\label{sec:introduction}
%what is the problem in general
\IEEEPARstart{I}{n} this paper we discuss the problem of error correcting codes with real valued entries. The goal is to recover an input vector \textbf{m} $\in \mathbb{R}^{k}$ from a corrupted measurement vector $\bf{y} = \textbf{Gm}+\bf{e}$, where $\textbf{G}\in \mathbb{R}^{n\times k}$ is a (coding) matrix that has full column rank ($n>k$ , $R:=\frac{k}{n}$ is the code rate) and $\bf{e} \in \mathbb{R}^n$ is a (sparse) error vector. If the vector $\bf{e}$ is known, then $\tilde{\bf{y}}=\bf{y}-\bf{e}=\bf{Gm}$ and since $\bf{G}$ has full column rank, $\bf{m}$ can be easily reconstructed from $\tilde{\bf{y}}$. Thus, reconstructing $\bf{m}$ from $\bf{y}$ is equivalent to reconstructing $\bf{e}$ from $\bf{y}$. By constructing a parity check matrix $\bf{H}$ \cite{Marshall_Coding_1984} which eliminates $\bf{G}$ (i.e. $\bf{HG}=\bf{0}$) we obtain the syndrome $\bf{s}$ which is defined as
\beq
\label{def:syndrome definition}
\vs=\textbf{H}\vy=\textbf{HGm}+\textbf{H}\ve=\textbf{H}\ve .
\eeq
Note that the syndrome $\bf{s}$ depends only on the error vector $\bf{e}$ and not on the input vector $\bf{m}$. Let $r=n-k$ be the redundancy of the code. Since $\bf{G}\in \mathbb{R}^{n\times k}$ is a full column rank matrix, its kernel has dimension $r$, thus $\bf{H}\in \mathbb{R}^{r\times n}$.

The sparsity requirement of $\bf{e}$ is intuitively explained by the fact that if the fraction of the corrupted entries is too large the reconstruction of $\bf{m}$ is impossible. Therefore, it is commonly assumed that only a few entries of $\bf{e}$ are non-zero
\beq
\left\|\bf{e} \right\|_{\ell_0}:=\left|\left\{i:e_i\neq 0\right\}\right|  \leq t(n).
\eeq
Given the coding matrix $\bf{G}$, it has been shown in \cite{Candes_Decoding_2005} that if $ t(n) > \frac{\rm{cospark}(G)}{2}$ it is impossible to recover $\bf{m}$ from $\bf{y}$, where the cospark of matrix $\bf{A}$ is defined as
\beq
\label{def:cospark}
\rm{cospark}(\textbf{A}) := \displaystyle\min_{\vx\in\mathbb{R}^k:x\neq 0}{\left\|\textbf{A}\vx \right\|_{\ell_0}}.
\eeq
This provides an upper bound on the number of errors that can be corrected. In a way, the cospark is the analog equivalent of the Hamming distance between the codewords. In \cite{Candes_Decoding_2005} it has been shown that $\rm{cospark}(\bf{G})=\rm{spark}(\bf{H})$, where the spark of a matrix is the minimal number of linearly dependent columns of a matrix:
\beq
\label{def:spark}
\rm{spark}(\textbf{H}) := \displaystyle\min_{\vx\in\mathbb{R}^n:x\neq 0}{\left\|\vx\right\|_{\ell_0}} \rm{\;\;subject\;\; to\;\;} \bf{Hx}=\bf{0}.
\eeq
From (\ref{def:spark}) it is easy to see that the largest number of correctable errors cannot exceed the rank of the parity check matrix. We assume that the error vector $\bf{e}$ is the sparsest vector that explains the input $\bf{y}$. Therefore, the decoding problem is reduced to finding a sparse solution to the underdetermined system:
\beq\
    \label{opt:p0problem}
    \displaystyle\min_{\bf{x}\in \mathbb{R}^{n}}{\left\|\bf{x}\right\|_{\ell_0}}  \rm{\;\;subject\;\; to\;\;}  \bf{Hx}=\bf{s}.
\eeq
This problem is NP-hard \cite{Natarajan_Sparse_1995}.

The performance of the code depends on the coding matrix $\bf{G}$ (or alternatively the parity check matrix $\bf{H}$) and the decoding technique. Wolf \cite{Wolf_Analog_1983} extracts $r=2t$ columns from the IDFT matrix and uses it as a coding matrix $\bf{G}$. Therefore, after the encoding we get a sequence of real (or complex) numbers $\tilde{\bf{y}}= \textbf{Gm}$, whose DFT has zeros in certain positions. He showed that the same technique for decoding BCH codes over finite field can be utilized to decode the real number code as well. He also showed that these decoding algorithms are tolerant to small errors at every entry in addition to the impulsive noise. Further work on these real BCH codes has been done by Henkel  \cite{Henkel_Multiple_1988}. He studied the influence of small additive noise on the detection of error locations by algebraic methods. In addition, he provided another proof of the main result of \cite{Wolf_Analog_1983} based on the Newton interpolation method. In this proof a different definition of syndrome is presented that make it possible to locate an error-free range of the codeword by observing this new syndrome (i.e. without any further operations).

There are many applications for analog coding. Gabay et al. \cite{Gabay_Real_2000} showed that a real BCH code can be used for simultaneous source coding and impulse noise cancellation. More specifically, they showed that simultaneously correcting the impulse noise and reducing the quantization noise by using BCH codes leads to a reduction in the end-to-end Peak Signal to Noise Ratio (PSNR). Henkel \cite{Henkel_Analog_2000} showed that using Wolf analog codes (also known as Analog Reed Solomon Codes) can reduce the high peak-to-average ratio (PAR) of multi carrier modulation signals. The clipping of the high peaks caused by analog circuitry leads to an impulsive additive noise. He showed that we can detect the positions of the noise impulses (by setting 95\% of the clipping amplitude $V_c$ as a threshold), and then Analog Reed Solomon (RS) erasure decoding can correct the clipping errors. Another source of impulsive noise on multi carrier modulation signals are nulls in the channel's frequency response. It is well known that uncoded orthogonal frequency-division multiplexing (OFDM) must cope with symbol recovery problems when the channel has nulls close to or on the FFT grid. Wang and Giannakis \cite{Wang_Complex_2003} introduced complex field precoding for OFDM (CFC-OFDM) where a complex-field coding is performed before the symbols are multiplexed to improve the average performance of OFDM over random frequency-selective channels. They provided design rules for achieving the maximum diversity order, and showed that if the channel is modeled with random taps, a good choice of the (analog) precoding matrix can enhance the average BER and suits any realization of the channel coefficients. In \cite{Henkel_OFDM_2005} Henkel and Hu showed that OFDM can be seen as an analog RS code if a cyclically consecutive range of carriers is not used for transmission, and in \cite{Abdelkefi_Improvement_2003} Abdelkefi et al. used the pilot tones of the OFDM system as a syndrome to correct impulsive noise in the presence of additive white Gaussian noise (AWGN). A different type of analog codes is presented in \cite{Scaglione_Optimal_2002}, \cite{Palomar_optimum_2004}. A linear space time block code is used to generate transmit redundancy over the real/complex field. However, these papers design optimal transmit redundancy for optimal Linear receivers and solve the coding problem under a MSE performance metric. Therefore, these coding designs are better for AWGN but not for impulsive noise.

Another related topic is Compressed Sensing (CS).  In Compressed Sensing we are given a representation dictionary $\bf{D}$ (defined as a compressed sensing matrix of size $r \times n$) and the rows of $\bf{D}$ are used to sample the information vector $\bf{x}$
\beq
\textbf{s}= \textbf{Dx} .
\eeq
Given the vector $\bf{s}$, which lies in the low dimensional space $\mathbb{R}^r$, we want to extract the information vector $\bf{x}$, which lies in the higher dimensional space $\mathbb{R}^n$. Under the assumption that the vector $\bf{s}$ is composed of as few columns of $\bf{D}$ as possible, we look for the sparsest vector $\tilde{\bf{x}}$ that explains $\bf{s}$. In other words, we are looking for the solution of equation (\ref{opt:p0problem}) with the replacement of $\bf{H}$ with $\bf{D}$. Under a certain condition on $\bf{H}$ and the size of the support of $\textbf{e}$, the sparsest solution of (\ref{opt:p0problem}) can be found by minimizing the $\ell_1$ norm instead of the $\ell_0$ norm \cite{Candes_Decoding_2005},\cite{Donoho_For_2004},\cite{Donoho_Optimally_2003},\cite{Elad_A_Generalized_2001}.
\beq\
    \label{opt:p1problem}
    \displaystyle\min_{\vx\in \mathbb{R}^{n}}{\left\|\vx\right\|_{\ell_1}}  \rm{\;\;subject\;\; to\;\;}  \textbf{H}\vx=\vs \;\;\;\;\;\; \rm{(P_1)}.
\eeq
The $\ell_1$ norm is convex and (\ref{opt:p1problem}) can be solved using Linear Programming (LP) \cite{Candes_Decoding_2005}.

Donoho and Elad in \cite{Elad_A_Generalized_2001} ,\cite{Donoho_Uncertainty_2001} and \cite{Elad_On_2001} introduced the term Incoherent Dictionary (or mutual incoherence property) which simply means that for every pair of columns of a dictionary $\mathbf{D}=\left[\bf{d}_1,\bf{d}_2,\ldots,\bf{d}_n\right]$
\beq
 \left|\left\langle \bf{d}_i,\bf{d}_j \right\rangle\right|\leq\frac{\mu}{\sqrt{r}}
\eeq
where $\mu$ is the coherence coefficient. They showed that in the special case where the CS matrices $\textbf{H}$ are constructed by concatenating two unitary matrices $\Phi$ and $\Psi$ of size $r \times r$, the equivalence between (\ref{opt:p0problem}) and (\ref{opt:p1problem}) holds for
$\left\|\vx\right\|_{\ell_0}\leq\frac{\sqrt{2}-0.5}{M}$, where $M$ is defined as
$M := \sup{\left\{\left|\left\langle \psi_i,\phi_j \right\rangle\right|, 1<i,j< r \right\}}$. In \cite{Donoho_Uncertainty_2001} it was shown that $1/\sqrt{r} \leq M \leq 1$. Hence, if the two unitary matrices $\Phi$ and $\Psi$ are chosen such that $M=1/\sqrt{r}$ (i.e. the coherence coefficient $\mu=1$) the equivalence holds as long as $\left\|\vx\right\|_{\ell_0}\leq\alpha n^{1/2}$ for some constant $\alpha$ (it was shown in  \cite{Donoho_For_2004} that $\alpha\approx0.65$).

Candes and Tao \cite{Candes_Decoding_2005} introduced the term Restricted Isometry Property (RIP), which measures how orthogonal the columns of $\bf{H}$ are. Let $\textbf{H}\in\mathbb{R}^{r\times n}$, $T$ be a subset of $\left\{1,2,...,n\right\}$ and $\bf{H}_T$ be a submatrix of $\bf{H}$, constructed by taking the columns of the matrix $\bf{H}$ indexed by $T$. The restricted isometry property of order $L$ is defined as the smallest number $\delta_L$ such that for all $\left|T\right|\leq L \:\:,\:\: \textbf{c} \in\mathbb{R}^{\left|T\right|}$
\beq\
    \label{def:RIP delta}
    \left(1-\delta_L\right)\left\|\textbf{c}\right\|^{2}_{\ell_2}\leq\left\|\bf{H}_T \textbf{c}\right\|^{2}_{\ell_2}\leq\left(1+\delta_L\right)\left\|\textbf{c}\right\|^{2}_{\ell_2}.
\eeq
It is easy to show (see \cite{Candes_Decoding_2005}) that if $\lambda\left(\bf{A}\right)$ is an eigenvalue of the matrix $\bf{A}$ then (\ref{def:RIP delta}) is equivalent to
\beq\
    \label{def:RIP eig}
  1-\delta_L\leq\lambda\left(\bf{H}^{T}_{T}\bf{H}_T\right)\leq 1+\delta_L  \:\:\: \forall \left|T\right|\leq L.
\eeq
The RIP is important since if the RIP constants satisfy
\beq\
    \label{ineq:good RIP}
    \delta_{t} + \delta_{2t} + \delta_{3t} < 1
\eeq
then problems (\ref{opt:p0problem}) and (\ref{opt:p1problem}) are equivalent when the size of the support of $\bf{e}$ is at most $t$. Therefore, if $\bf{H}$ has a "good" RIP one can correct any $t$ errors using Linear Programming. The Gaussian random matrices satisfy (\ref{ineq:good RIP}) for $\left\|e \right\|_{\ell_0} \leq\rho n$ ($\rho<<1$) with a probability of $1-\epsilon(n)$ where $\epsilon(n)$ decays exponentially to zero with $n$.

In some applications, deterministic matrices with the Restricted Isometry Property are desirable due to storage limitations. However, there is no deterministic construction of a matrix for which (\ref{ineq:good RIP}) holds with a constant fraction of the block length. DeVore \cite{DeVore_Deterministic_2007} used a polynomial construction to obtain a matrix that satisfies RIP for $\left\|e \right\|_{\ell_0} \leq \alpha n^{\frac{1}{m+1}}$ where $m>1$ is an arbitrary integer. The matrices of \cite{DeVore_Deterministic_2007} provide a higher code rate than the matrices in Donoho and Elad \cite{Elad_A_Generalized_2001}-\cite{Donoho_Uncertainty_2001}, which have a rate of $R=\frac{k}{n}=0.5$, but DeVore's matrices correct fewer than $\alpha\sqrt{n}$ errors. Note that when the RIP fails, there is no guarantee that the $\ell_1$ minimization (\ref{opt:p1problem}) will compute the sparsest solution. Unfortunately, verifying the RIP for a given matrix $\bf{H}$ is a difficult task with exponential complexity. This property requires checking (\ref{def:RIP delta}) for all sub matrices of selecting $t$ arbitrary columns. Lee and Bresler \cite{Lee_Computing_2008} used the $\ell_1$ relaxation and some additional relaxations to verify the RIP in polynomial time by using Semidefinite Programming (SDP). Statistical versions of the RIP (STRIP for short) were introduced by Gurevich et al. \cite{Gurevich_The_2009} and by Calderbank et al. \cite{Calderbank_Construction_2009}. Both versions bound the probability that the RIP holds for an L-sparse \textbf{random} vector (i.e. the L entries of the vector chosen at random). \cite{Gurevich_The_2009} showed that the STRIP holds in general for any incoherent matrix. In \cite{Calderbank_Construction_2009} Calderbank et al. bound STRIP's performance for a large class of deterministic complex matrices. More specifically, they showed that under the assumption that the matrix $\bf{H}\in\mathbb{C}^{r\times n}$ has columns that form a group under point-wise multiplication and rows that are orthogonal and vanish under summation (the row sums are equal to zero), the RIP (\ref{def:RIP delta}) holds for $1>\delta_L>\frac{L-1}{n-1}$ for any L-sparse random vector $\bf{x}$ with a probability of
\beq
\label{P STRIP}
\rm{P_{RIP}}(\bf{x})=1-\frac{\frac{2L}{r}+\frac{2L+7}{n-3}}{\left(\delta_L-\frac{L-1}{N-1}\right)^2}.
\eeq
It was pointed out in \cite{Gan_Analysis_2009} that this assumption is too weak since almost all linear codes meet these conditions (for example a partial DFT matrix when excluding the first row), however it is not guaranteed that they will perform well for compressed sensing or equivalently for decoding linear analog correcting codes by the $\ell_1$ minimization (\ref{opt:p1problem}). In \cite{Gan_Analysis_2009} Gan et al. showed a tighter bound on the performance of the STRIP in the case of matrices that nearly meet the Welch bound (which is a stronger restriction on the dictionary that bounds the mutual coherence of the matrix $\mathbf{H}$ - $\displaystyle\max_{i\neq j}{\left|\left\langle \bf{h}_i,\bf{h}_j \right\rangle\right|}$). It has been shown that for these matrices the RIP holds with probability that exponentially decays with (r/L).

In \cite{Near_Berrou_1993} Turbo codes were first introduced. Their performances in terms of bit error rate (BER) are close to the Shannon limit. In \cite{Near_Pyndiah_1998} a coding scheme of block turbo codes (BTC) was described, where two (or more) encoders are serially concatenated to perform a product code. The product codes are used in the area of digital error correction codes (i.e., codes over a finite field) and are very efficient for building long block codes by using several short blocks. The decoding of such codes can be done sequentially, using one decoder at a time. In \cite{Mura_Iterative_2003},\cite{Hu_Turbo_2005} analog products codes are presented. The $N^2$-length information sequence is reshaped into an $N\times N$ information matrix. Then, the encoding is done by adding a parity check component to each columns and row of the information matrix such that the columns and rows of the $(N+1)\times(N+1)$ encoded matrix are sum to zero. This process describes a product code with analog parity check component. The decoding was done using an iterative decoder that converges to the least squares solution. However, in contrast to the method described on this paper, these product codes are optimized for MSE criterion instead of $\ell_1$ criterion required
for sparse reconstruction.

As with the statistical version of the RIP, in this paper we weaken the strong RIP constant requirement at the price of an arbitrary small probability of error. However, in contrast to STRIP, in this effort the problem of decoding the long block code is decoupled into two sets of parallel Linear Programming problems, which leads to much lower complexity than solving (\ref{opt:p1problem}) to decode the codeword at once (see \ref{sec:decoding with probability}). In other words, the reconstruction of $\bf{e}$ from $\bf{y}$ is performed using LP (iteratively) even though $\left\|\bf{e}\right\|_{\ell_0}$ is higher than what is required by the RIP, with the caveat that for a few ensembles of errors the reconstruction fails. More specifically, inspired by the iterative decoding of Turbo block codes \cite{Near_Pyndiah_1998} we show that given a code capable of correcting $\alpha \sqrt{n}$ errors, we can construct a turbo analog block code that is capable of correcting up to $\frac{\alpha n^{3/4}}{log n}$ errors with a probability of $1-\epsilon(n)$ where $\epsilon(n)$ decays sub exponentially to zero. This provides a simple analog coding procedure that improves existing deterministic coding matrices by using a probabilistic approach.

The outline of the paper is as follows. Section \ref{sec:problem formulation} describes the analog product code and a mathematical formulation of the problem. Section \ref{sec:decoding with probability} gives the solution and a bound on the probability of decoding and the complexity of this turbo analog decoder. Section \ref{sec:simulations} provides simulation results for the extended Donoho matrices described  in section \ref{sec:decoding with probability}. We end up with some conclusions.
%%%%%%%%%%%%%%%%%%%%%%%%%%%%%%%%%%%%%%%%%%%%%%%%%%%%%%%%%%%%%%%%%%%%%%%%%%%%%%%%%%%%%%%%%%%%%%%%%%%%%%%%%%%%%%
%%%%%%%%%%%%%%%%%%%%%%%%%%%%%%%%%%%%%%%%%%%%%%%%%%%%%%%%%%%%%%%%%%%%%%%%%%%%%%%%%%%%%%%%%%%%%%%%%%%%%%%%%%%%%%
%%%%%%%%%%%%%%%%%%%%%%%%%%%%%%%%%%%%%%%%%%%%%%%%%%%%%%%%%%%%%%%%%%%%%%%%%%%%%%%%%%%%%%%%%%%%%%%%%%%%%%%%%%%%%%
%%%%%%%%%%%%%%%%%%%%%%%%%%%%%%%%%%%%%%%%%%%%%%%%%%%%%%%%%%%%%%%%%%%%%%%%%%%%%%%%%%%%%%%%%%%%%%%%%%%%%%%%%%%%%%

\section{analog product codes and problem formulation}
\label{sec:problem formulation}
Suppose that we want to encode a vector $\textbf{m}\in\mathbb{R}^k$, where $k=K^2$ $K\in N$. Suppose that we reshape the vector into a matrix $\textbf{M}\in \mathbb{R}^{K \times K}$. Assume we are given a code generator matrix $\bf{G}\in \mathbb{R}^{N \times K}$. Let $R_i=\frac{K}{N}$ be the code rate of $\bf{G}$. The analog product coding process is as follows:
\begin{enumerate}
    \item inner code - code each column of $\bf{M}$ using the coding matrix $\bf{G}$ to produce a new matrix $\tilde{\bf{M}}\in\mathbb{R}^{N\times K}$.
    \item outer code - code each row of $\tilde{\bf{M}}$ using the coding matrix $\bf{G}$ to produce a new matrix $\tilde{\bf{Y}}\in\mathbb{R}^{N \times N}$.
\end{enumerate}
Let $R=\frac{k}{n}=R_{i}^{2}$ be the code rate of the analog product code, where $n=N^2$. This process can be written more compactly as
\beq
    \label{eq:compact way to encode matrix}
    \tilde{\bf{Y}}=\tilde{\bf{M}}\bf{G^T}=\bf{GMG}^{T},
\eeq
where $\bf{G}^{T}$ is $\bf{G}$ transpose. It easy to see from (\ref{eq:compact way to encode matrix}) that the order of the two stages above is irrelevant. As in section \ref{sec:introduction}, we assume that the model is $\bf{Y}=\tilde{\bf{Y}}+\bf{E}$. Therefore, $\bf{Y}=\bf{GMG}^{T}+\bf{E}$, where $\bf{E}\in \mathbb{R}^{\sqrt{n} \times \sqrt{n}}$ is the arbitrary (sparse) error vector presented as a matrix. Since $\bf{G}$ has full rank, decoding $\bf{M}$ from $\bf{Y}$ is equivalent to reconstructing $\bf{E}$ from $\bf{Y}$. By the linearity of the code, the parity check matrix $\bf{H}\in\mathbb{R}^{N-K\times N}$ such that $\bf{HG}=\bf{0}$ provides a set of equations that do not depend on the input matrix $\bf{M}$.
\begin{eqnarray}
  \label{eqa:reconstruct E from HYT HY}
    \nonumber \bf{HY}=\bf{HGMG}^{T}+\bf{HE}=\bf{HE}\\
    \bf{YH}^{T}=\bf{GMG}^{T}\bf{H}^{T}+\bf{EH}^{T}=\bf{EH}^{T}
\end{eqnarray}
Denote $\left\|A\right\|_{\ell_0}:=\left|\left\{\left(i,j\right):A_{ij}\neq0\right\}\right|$. The decoding problem becomes
\beq\
    \label{eq:reconstruct E from HYYT}
    \displaystyle\min_{\bf{E}\in\mathbb{R}^{N\times N}}{\left\|\bf{E}\right\|_{\ell_0}} \:\: s.t \:\: \bf{H}(\bf{Y}|\bf{Y}^{T})=\bf{H}(\bf{E}|\bf{E}^{T}).
\eeq
%%%%%%%%%%%%%%%%%%%%%%%%%%%%%%%%%%%%%%%%%%%%%%%%%%%%%%%%%%%%%%%%%%%%%%%%%%%%%%%%%%%%%%%%%%%%%%%%%%%%%%%%%%%%%%
%%%%%%%%%%%%%%%%%%%%%%%%%%%%%%%%%%%%%%%%%%%%%%%%%%%%%%%%%%%%%%%%%%%%%%%%%%%%%%%%%%%%%%%%%%%%%%%%%%%%%%%%%%%%%%
%%%%%%%%%%%%%%%%%%%%%%%%%%%%%%%%%%%%%%%%%%%%%%%%%%%%%%%%%%%%%%%%%%%%%%%%%%%%%%%%%%%%%%%%%%%%%%%%%%%%%%%%%%%%%%
%%%%%%%%%%%%%%%%%%%%%%%%%%%%%%%%%%%%%%%%%%%%%%%%%%%%%%%%%%%%%%%%%%%%%%%%%%%%%%%%%%%%%%%%%%%%%%%%%%%%%%%%%%%%%%
\section{The probability of error for the two step iterative LP Turbo decoder}
\label{sec:decoding with probability}
In this section we show that any code that is capable of correcting up to $\alpha\sqrt{n}$ errors can be extended by the scheme of Turbo codes to a code correcting up to $\frac{\alpha n^{3/4}}{log n}$ with a probability of error going to zero as a function of $n$. Let $\bf{G}$ be a generator matrix of a code that is capable of correcting up to $\alpha\sqrt{n}$ errors. Let the coding process be as shown in section \ref{sec:problem formulation}. The main theorem is that if $\left\|\bf{E}\right\|_{\ell_0}\leq\frac{\alpha n^{3/4}}{log n}$, one can find the solution to (\ref{eq:reconstruct E from HYYT}) with a probability approaching $1$ as a function of $n$.

To prove the above, we use a two-step decoding procedure. First we decode each row of $\bf{Y}$ independently using (\ref{opt:p1problem}) and correct the errors found in this step, then decode each column of the corrected matrix in the same way. Then, we bound the probability of error of the two-step decoder and show that the bound decays to zero (sub exponentially) as the block size increases.

For the decoding process we use the following notation, given a matrix $\bf{A}$. We denote the $j$'th row of $\bf{A}$ by $(\bf{A}^T)_j$ and the $i$'th column of $\bf{A}$ by $\bf{A}_i$. The decoding process of the outer code is as follows. Let $\hat{\bf{E}}$ be the error of the outer code. Each row of $\hat{\bf{E}}$ can be found by solving (\ref{opt:p1problem}) for each row of $\bf{Y}$ sequentially:
\beq\
\label{eq:decoding rows}
(\hat{\bf{E}}^T)_j= arg\displaystyle\min_{\textbf{x}_j\in\mathbb{R}^N}{\left\|\textbf{x}_i\right\|_{\ell_1}} \:\: s.t \:\: \bf{H}(\textbf{Y}^{T})_{i}=\textbf{Hx}_i \;\;i=1,...,N.
\eeq
Following the notation of (\ref{eq:compact way to encode matrix}), this gives us $\hat{\tilde{\bf{M}}}$, the decoded matrix of the outer code, $\hat{\tilde{\bf{M}}}\in \mathbb{R}^{N \times K}$.

The decoding process of the inner code is done as follows. Let $\check{\bf{E}}\in \mathbb{R}^{N\times K}$ be the error of the inner code, $\hat{\tilde{\bf{M}}}=\bf{GM}+\check{\bf{E}}$. Each column of $\check{\bf{E}}$ can be found by decoding each column of $\hat{\tilde{\bf{M}}}$ sequentially.
\beq
\label{eq:decoding columns}
\check{\bf{E}}_i:=arg\displaystyle\min_{\textbf{x}_i\in\mathbb{R}^N}{\left\|\textbf{x}_i\right\|_{\ell_1}} \:\:
s.t \:
\bf{H}(\textbf{Y}-\hat{\textbf{E}})_i=\bf{H}\textbf{x}_i
\;\;i=1,...,K.
\eeq
The main theorem is that this two-step decoder correctly decodes the codeword and gives the sparsest solution of (\ref{eq:reconstruct E from HYYT}) with a probability approaching one sub exponentially with $n$.

Moreover, the problem of decoding the long block code is decoupled into two sets of parallel Linear Programming problems. This decoupling leads to a lower complexity than solving (\ref{opt:p1problem}) to decode the codeword at once. More specifically, decoding a long codeword with size $n$ using Linear Programming as in (\ref{opt:p1problem}) takes $O(n^{3.5})$  operations \cite{Nemirovski_Optimization_1999}. For the outer decoder, each row is decoded using LP with $O(N^{3.5})$ operations; there are $N$ such rows. For the inner decoder each column is decoded with $O(N^{3.5})$; there are $K$ such columns. Using the relation $n=N^2$ and assuming $N\approx K$ the iterative decoder decodes with only $O(n^{2.25})$ operations.

Again from (\ref{eq:compact way to encode matrix}) it is easy to see that the decoding procedure can be done in the reverse order; i.e. first decode column by column and then row by row. Because the constraints are independent the decoding procedure can be rewritten as
\beq\
\label{opt:decoding rows matrix form}
\hat{\bf{E}}:=arg\displaystyle\min_{\textbf{X}\in\mathbb{R}^{N\times N}}{\displaystyle\sum_{i,j}{\left|x_{i,j}\right|}} \:\:
s.t \:\:\:\:\:\:\:
\bf{HY}^{T}=\bf{HX}^{T}
\eeq
\beq\
\label{opt:decoding columns matrix form}
\check{\bf{E}}:=arg\displaystyle\min_{\textbf{B}\in\mathbb{R}^{N\times K}}{\displaystyle\sum_{i,j}{\left|b_{i,j}\right|}} \:\:
s.t \:\:\:\:\:\:\:
\bf{H\hat{\tilde{M}}}=\bf{HB}
\eeq
where $\bf{\hat{\tilde{M}}}$ can be found by solving
\beq
\label{eq:decoding rows using E_hat}
\bf{\hat{\tilde{M}}G^T}=\bf{Y}-\hat{\bf{E}}
\eeq

To intuit why this two step decoder leads to the solution of (\ref{eq:reconstruct E from HYYT}), consider a scenario in which only the first row has more than $\alpha N^{1/2} =\alpha n^{1/4}$ errors. Assume the worst case is that if a codeword is decoded erroneously every entry of the word is wrong. After we decode row by row as in (\ref{eq:decoding rows}), every row except the first one will be error free (since the code is capable of decoding up to $\alpha N^{1/2}$ errors). Thus $N$ errors shift to the inner code such that there is only a single error in each column. This can be corrected by decoding the columns as in (\ref{eq:decoding columns}).

One should bear in mind that if the number of errors is bounded by $t(n)$, the worst case for the two-step decoder is that there is no row that is completely filled with errors. Suppose that the total number of errors on the block is $t(n)$ and a certain row has $t_1>\alpha n^{1/4}$ errors; without loss of generality assume it is the first row. Thus, the rest of the block has $t(n)-t_1$ errors. After decoding row by row, we assume that the first row is decoded with errors no matter how large $t_1$ is (because $t_1>\alpha n^{1/4}$). Therefore, for larger $t_1$, there are fewer errors left for the rest of the block and it has a higher probability of being decoded without errors.

\lemma{Given a code that is capable of correcting $\alpha \sqrt{N}$ errors, the decoding procedure described by (\ref{opt:decoding rows matrix form})-(\ref{eq:decoding rows using E_hat}) provides a complete burst protection for bursts with sizes up to $t_b(n):=\alpha n^{3/4}-n^{1/2}+2\alpha n^{1/4}+1$ for any block size $n=N^2$ (under the assumption that there are no other errors on the decoded block).}

\proof{Assume the vector $\vy$ corrupted by $t_b(n)$ consecutive errors. Since $t_b(n)=n^{1/2}(\alpha n^{1/4}-1)+2\alpha n^{1/4}+1$, reshaping the vector $\vy$ into a matrix with size $n^{1/2}\times n^{1/2}$ causes there to be $(\alpha n^{1/4}-1)$ rows that are completely filled with errors, and two other rows that together have $2\alpha n^{1/4}+1$ errors. After decoding the outer code as in (\ref{opt:decoding rows matrix form}) there will be no more than $\alpha n^{1/4}$ rows with errors. In other words, there will be no more than $\alpha n^{1/4}$ in each column. Therefore, the inner decoder (\ref{opt:decoding columns matrix form}) correct all the errors, and we decode the block correctly.} \endproof

\theorem{Let $\bf{G}\in \mathbb{R}^{K\times N}$ be a generator matrix of a code that is capable of correcting $\alpha \sqrt{N}$ errors, and let $n=N^2$. The two-step decoding procedure described by (\ref{opt:decoding rows matrix form})-(\ref{eq:decoding rows using E_hat}) provides a turbo analog block code that is capable of correcting up to $t(n):=\frac{\alpha n^{3/4}}{log n}$ errors with a probability of $1-\epsilon(n)$, where $\epsilon(n)$ decays sub exponentially to zero with $n$. }

\proof{The code fails to recover the correct word if the number of codewords that are decoded with errors in the outer code is higher than $\alpha n^{1/4}$. In other words, if there are more than $\alpha n^{1/4}$ rows with more than $\alpha n^{1/4}$ errors, the code will fail to recover the correct word. We want to bound the probability of that event. By assumption there are $t(n)$ errors and $N=\sqrt{n}$ rows. Set a random i.i.d binary process $x_i$ with
\beq
\label{probability to choose row}
p:=P(x_i=1)=\frac{1}{N}=\frac{1}{\sqrt{n}}\:\:,\:\:\:i=1,2,...,t(n).
\eeq
Let $y=\displaystyle\sum_{i=1}^{t(n)}{x_i}$ be a binomial random variable with probability $p$. This is expressed as,
\beq
\label{y binomial distribution}
y \sim  B(t(n),p).
\eeq
Therefore, the probability that a given row will have more than $\alpha n^{1/4}$ errors can be bounded by the Chernoff bound.
\beq
\label{ineq:chernoff bound on the row probability of error}
P(y>\alpha n^{1/4})\leq e^{-s\alpha n^{1/4}}\left(p e^{s}+1-p\right)^{t(n)}.
\eeq
Taking the derivative of the RHS of (\ref{ineq:chernoff bound on the row probability of error}) and equating to zero leads to
\beq
s=\log\left(\frac{(1-p)\alpha n^{1/4}}{p(t(n)-\alpha n^{1/4})}\right).
\eeq
Where $s>0$ if
\beq
\alpha n^{1/4}<t(n)<\alpha n^{3/4}.
\eeq
Choosing
\beq
\label{eq:chosen t_n}
t(n)=\frac{\alpha n^{3/4}}{\log(n)},
\eeq
it is shown in appendix \ref{subsec:Chernoff bound of the outer code} that for all $n\geq2$
\beq
\label{ineq:Chernoff bound for row error}
p(y>\alpha n^{1/4})\leq q(n)
\eeq
where
\beq
q(n):=\left(\frac{1-n^{-1/2}}{\frac{1}{log(n)} -n^{-1/2}}\right)^{\frac{\alpha n^{3/4}}{log(n)}-\alpha n^{1/4}}\left(log(n)\right)^{\frac{-\alpha n^{3/4}}{log(n)}}.
\eeq
Further simplification yields:
\beq
\label{ineq:Chernoff bound for row error bound}
q(n)\leq e^{-\alpha n^{1/4}(\log\log(n)-\frac{3}{2}+\frac{1}{log(n)}-\frac{1}{\sqrt{n}})}.
\eeq
Therefore, since the total number of errors is bounded by $t(n)$, we can uniformly bound the probability that a given row will be decoded with
errors by $q(n)$. For the inner code, we want to bound the probability that more than $\alpha n ^{1/4} $ rows are decoded with
errors. Assume the worst case that if a row is decoded with an error then all elements in the row are wrong. Denote the number of rows with errors by
$\tilde{Z}$. We uniformly bound the probability that a given row is decoded with errors by $q(n)$ . Define a random i.i.d binary
process $y_i$ with $P(y_i=1)=q$ , $i=1,2,...,n^{1/2}$, set $Z=\displaystyle\sum_{i=1}^{n^{1/2}}{y_i}$ a binomial random variable
with a probability of $p=q(n)$. Therefore, $P(\tilde{Z}>\alpha n^{1/4})\leq P(Z>\alpha n^{1/4})$ which can be bounded by the Chernoff bound.
Choosing
\beq
s=\log\left(\frac{(1-q)\alpha n^{1/4}}{q(n^{1/2}-\alpha n^{1/4})} \right).
\eeq
Since $\alpha\leq 1$, there exists a number $N_0$ (typically a small number) such that for all $n\geq N_0$ we have $s > 0$.
A simple computation yields (see appendix \ref{subsec:Chernoff bound of the inner code}):
\beq
\label{ineq:Chernoff bound for block error}
\mbox{P(block error)}\leq P_b
\eeq
where
\beq
\label{eq:p_b}
P_b:=e^{-\alpha^2 n^{1/2}(\log\log(n)-\frac{3}{2}+\frac{\log(1-\alpha n^{-1/4})}{\alpha^2}-\frac{n^{-	1/4}\log(\frac{n^{1/4}}{\alpha}-1)}{\alpha})}.
\eeq
This bound decays sub-exponentially in the block size n. Therefore, the two-step decoder described by (\ref{opt:decoding rows matrix form})-(\ref{eq:decoding rows using E_hat}) finds the sparsest solution of (\ref{eq:reconstruct E from HYYT}) with a probability of error decaying to zero as in (\ref{eq:p_b}) when using Linear Programing.}
\endproof

\begin{figure}[htbp]
    \centering
        \includegraphics[width=0.4\textwidth]{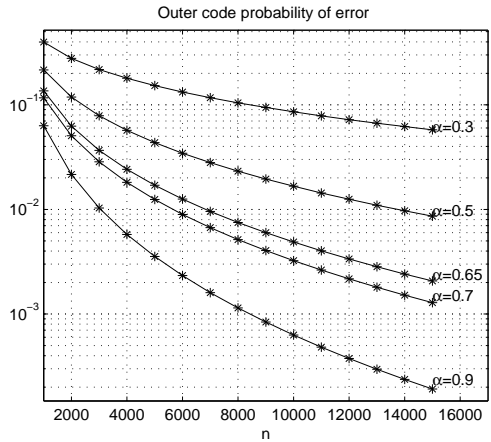}
    \label{fig:row_probability_bound}
    \caption{Upper bound on the outer code probability of error, equation (\ref{ineq:Chernoff bound for row error bound}) }
\end{figure}
\begin{figure}[htbp]
    \centering
        \includegraphics[width=0.4\textwidth]{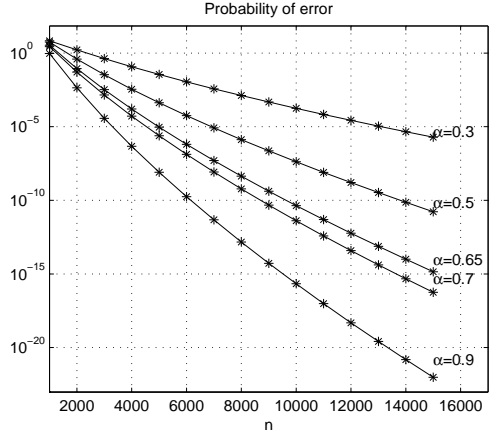}
    \label{fig:block_probability_bound}
    \caption{Upper bound on the probability of error, equation (\ref{ineq:Chernoff bound for block error})}
\end{figure}
%%%%%%%%%%%%%%%%%%%%%%%%%%%%%%%%%%%%%%%%%%%%%%%%%%%%%%%%%%%%%%%%%%%%%%%%%%%%%%%%%%%%%%%%%%%%%%%%%%%%%%%%%%%%%%
%%%%%%%%%%%%%%%%%%%%%%%%%%%%%%%%%%%%%%%%%%%%%%%%%%%%%%%%%%%%%%%%%%%%%%%%%%%%%%%%%%%%%%%%%%%%%%%%%%%%%%%%%%%%%%
%%%%%%%%%%%%%%%%%%%%%%%%%%%%%%%%%%%%%%%%%%%%%%%%%%%%%%%%%%%%%%%%%%%%%%%%%%%%%%%%%%%%%%%%%%%%%%%%%%%%%%%%%%%%%%
%%%%%%%%%%%%%%%%%%%%%%%%%%%%%%%%%%%%%%%%%%%%%%%%%%%%%%%%%%%%%%%%%%%%%%%%%%%%%%%%%%%%%%%%%%%%%%%%%%%%%%%%%%%%%%

\section{Numerical Experiments}
\label{sec:simulations}
In this section we investigate the performance of the two-step decoder in two sets of simulations. In the proof of the main theorem we uniformly bound the probability that a given row is decoded with errors. Therefore, in the first set of simulations we check the tightness of this bound. The number of errors $t(n)=\alpha\frac{n^{3/4}}{\log{n}}$ is fixed and we check how frequently a "bad" ensemble of errors has been chosen in vectors with a support size of $t(n)$ selected at random, for various sizes of block $n$. By a "bad" ensemble we mean an ensemble of errors that has more than $\alpha n^{1/4}$ rows with more than $\alpha n^{1/4}$ errors. The results are shown in table \ref{tab:uniform bound of block error} for $\alpha=0.65$.
\begin{table}[htbp]
	\centering
		\caption{"bad" ensemble frequency for $t(n)=\frac{\alpha n^{3/4}}{\log{n}}$}
		\label{tab:uniform bound of block error}
		\begin{tabular}{|p{1cm}|c|c|}\hline
				$n$ & $\log_{10}({P_{\rm{bad \; ensemble}}})$ & {number of errors - t(n)} \\\hline
				$81$ & $-1.6$ & $4$  \\\hline
				$441$ & $-4.21$ & $10$  \\\hline
				$1369$ & $-8.88$ & $20$  \\\hline
				$3481$ & $<-11$ & $36$  \\\hline
		\end{tabular}
\end{table}

\begin{figure}[htbp]
    \centering
        \includegraphics[width=0.45\textwidth]{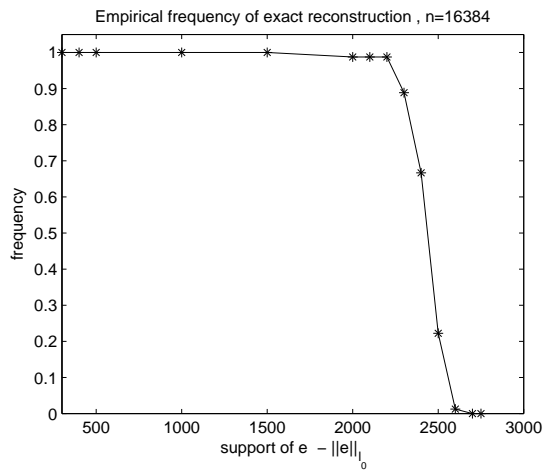}
    \label{fig:simulate donoho eye hadamard 128 criterion max error 15E-7}
    \caption{Reconstruction frequency of E with a support size of $\left\|e\right\|_{\ell_0}$ from $H(E|E^T)$ using iterative LP for decoding}
\end{figure}

In the second set of simulations we simulated the analog turbo block decoder that was shown in section \ref{sec:decoding with probability}, to recover $\bf{M}$ from $\bf{Y}=\bf{GMG}^T+\bf{E}$:

\begin{enumerate}
    \item $N=128$.
    \item In the simulation we used Donoho matrix composed of an Identity matrix and a Hadamard matrix of size $N/2$ each.
    \item Take the support set of size $t$ uniformly at random, and sample a vector $\ve$ at size $n=N^2$ with i.i.d Gaussian entries on the selected support.
    \item Reshape $\bf{e}$ to a structure of matrix $\bf{E}$ of size $N\times N$ .
    \item Put $\bf{Y}$=$\bf{E}$ (equivalent to choosing $\bf{M}=\bf{0}$  , there is no loss of generality since the code is linear).
    \item Reconstruct $\tilde{\bf{E}}$ from $\bf{Y}$ by solving equations (\ref{opt:decoding rows matrix form})-(\ref{eq:decoding rows using E_hat}).
    \item Compare $\tilde{\bf{E}}$ to $\bf{E}$.
    \item Repeat for various sizes of $t$ ( 240 times for each $t$).
\end{enumerate}
The results are presented in figure \ref{fig:simulate donoho eye hadamard 128 criterion max error 15E-7}. Our experiment shows that the input is recovered all the time as long as $\left\|\bf{e}\right\|_{\ell_0}\leq1500$. Note that we prove that we correctly reconstruct $\bf{E}$ as long as $\left\|\bf{e}\right\|_{\ell_0}\leq97$  (put $n$ and $\alpha=0.65$ as was shown in \cite{Donoho_For_2004} for Donoho matrices). In other words, the simulation results show that the actual performance of the Turbo analog scheme is much better than what has been proven. One explanation for this discrepancy is that Donoho's construction has been proven to guarantee correction as long as there are no more than $\alpha n^{1/2}$ errors, but some ensembles of errors can be corrected even though there are more errors than have been proven. A second explanation is that the uniform bound in the main theorem is very loose, as can been seen from table \ref{tab:uniform bound of block error}. The third explanation is that in the proof of the main theorem we chose $\frac{\alpha n^{3/4}}{\log{n}}$ as the number of errors,	 but it can easily be shown that one can select $\frac{n^{3/4}}{f(n)}$ and get a similar bound on the probability of error, where $f(n)$ is a monotonically increasing function for all $n>n_0$ (for some large enough $n_0$). However, increasing the number of errors leads to a slower decay of the probability of error (see table \ref{tab:uniform bound of block error for loglogn} for the example of $f(n)=\log{\log{n}}$).

\begin{table}[htbp]
	\centering
		\caption{"bad" ensemble frequency for $t(n)=\frac{\alpha n^{3/4}}{\log{\log{n}}}$}
	\label{tab:uniform bound of block error for loglogn}
		\begin{tabular}{|p{1cm}|c|c|}\hline
				$n$ & $\log_{10}({P_{\rm{bad \; ensemble}}})$ & {number of errors - t(n)}\\\hline
				$441$ & $-1.67$ & $34$  \\\hline
				$1369$ & $-2.13$ & $73$  \\\hline
				$3481$ & $-2.6$ & $140$ \\\hline
				$7225$ & $-3.5$ & $233$  \\\hline
				$13225$ & $-4.8$ & $356$  \\\hline
		\end{tabular}
\end{table}

%%%%%%%%%%%%%%%%%%%%%%%%%%%%%%%%%%%%%%%%%%%%%%%%%%%%%%%%%%%%%%%%%%%%%%%%%%%%%%%%%%%%%%%%%%%%%%%%%%%%%%%%%%%%%%
%%%%%%%%%%%%%%%%%%%%%%%%%%%%%%%%%%%%%%%%%%%%%%%%%%%%%%%%%%%%%%%%%%%%%%%%%%%%%%%%%%%%%%%%%%%%%%%%%%%%%%%%%%%%%%
%%%%%%%%%%%%%%%%%%%%%%%%%%%%%%%%%%%%%%%%%%%%%%%%%%%%%%%%%%%%%%%%%%%%%%%%%%%%%%%%%%%%%%%%%%%%%%%%%%%%%%%%%%%%%%
%%%%%%%%%%%%%%%%%%%%%%%%%%%%%%%%%%%%%%%%%%%%%%%%%%%%%%%%%%%%%%%%%%%%%%%%%%%%%%%%%%%%%%%%%%%%%%%%%%%%%%%%%%%%%%
\section{conclusion}
\label{sec:conclusion}
In this paper we have presented a simple analog coding procedure that improves existing deterministic coding matrices by using a probabilistic approach. The proposed coding/decoding scheme is able to correct up to $\frac{\alpha n^{3/4}}{\log n}$ errors by solving a set of LP problems iteratively. This scheme shows a significant reduction in decoding complexity as compared to one-step LP decoding. Here we weakened the RIP requirement by allowing a vanishingly small probability of error where a Chernoff bound on the probability of error shows a sub-exponential decay to zero with the increase in block size. Moreover, simulation results show much better performance by this scheme.
%%%%%%%%%%%%%%%%%%%%%%%%%%%%%%%%%%%%%%%%%%%%%%%%%%%%%%%%%%%%%%%%%%%%%%%%%%%%%%%%%%%%%%%%%%%%%%%%%%%%%%%%%%%%%%
%%%%%%%%%%%%%%%%%%%%%%%%%%%%%%%%%%%%%%%%%%%%%%%%%%%%%%%%%%%%%%%%%%%%%%%%%%%%%%%%%%%%%%%%%%%%%%%%%%%%%%%%%%%%%%
%%%%%%%%%%%%%%%%%%%%%%%%%%%%%%%%%%%%%%%%%%%%%%%%%%%%%%%%%%%%%%%%%%%%%%%%%%%%%%%%%%%%%%%%%%%%%%%%%%%%%%%%%%%%%%
%%%%%%%%%%%%%%%%%%%%%%%%%%%%%%%%%%%%%%%%%%%%%%%%%%%%%%%%%%%%%%%%%%%%%%%%%%%%%%%%%%%%%%%%%%%%%%%%%%%%%%%%%%%%%%
\appendices
\section{Chernoff bound on the probability of error of the outer code}
\label{subsec:Chernoff bound of the outer code}
The probability that a given row will be erroneously decoded is bounded using the Chernoff bound. Let y be as in (\ref{y binomial distribution}) i.e. $y \sim B(t(n),p)$ where $p=n^{-1/2}$. Assume $n\geq2$
\beq
\bea{l}
\label{ineq:chernoff bound on the row probability of error appendix}
P\left(y>\alpha n^{1/4})\right)\leq e^{-s\alpha n ^{1/4}}E\left\{e^{sy}\right\}=\;\;\;\;\;\;\;\; \forall s>0. \\
=e^{-s\alpha n^{1/4}}\left(p e^s+1-p\right)^{t(n)}\\
\ena
\eeq
Let
\beq
\label{opt:optimize s of Chernoff bound}
s=arg\displaystyle\min_{s>0}{\:\:\:\:e^{-s\alpha n^{1/4}}\left(p e^s+1-p\right)^{t(n)}}.
\eeq
Taking the derivative of the RHS of (\ref{ineq:chernoff bound on the row probability of error appendix}) and equating to zero leads to
\beq
s=\log\left(\frac{(1-p)\alpha n^{1/4}}{p(t(n)-\alpha n^{1/4})}\right).
\eeq
Where $s>0$ if
\beq
\alpha n^{1/4}<t(n)<\alpha n^{3/4}\\
\eeq
Choosing
\beq
t(n)=\frac{\alpha n^{3/4}}{\log(n)}
\eeq
get
\beq
\bea{l}
p(y>\alpha n^{1/4})\leq q(n)\\
q(n):=\frac
%mone
{\left(1-n^{-1/2}\right)^{-\alpha n^{1/4}+\frac{\alpha n^{3/4}}{log(n)}}\left(\alpha n ^{1/4}\right)^{-\alpha n^{1/4}}\left(\frac{\alpha n^{3/4}}{log(n)}\right)^{\frac{\alpha n^{3/4}}{log(n)}}}
%mechane
{\left(n^{-1/2}\right)^{-\alpha n^{1/4}}\left(\frac{\alpha n^{3/4}}{log(n)}-\alpha n^{1/4}\right)^{-\alpha n^{1/4}+\frac{\alpha n^{3/4}}{log(n)}}}=\\

\frac
{%mone
\left(1-n^{-1/2}\right)^{-\alpha n^{1/4}+\frac{\alpha n^{3/4}}{log(n)}}\left(\alpha n^{3/4}\right)^{\frac{\alpha n^{3/4}}{log(n)}}\left(log(n)\right)^{-\frac{\alpha n^{3/4}}{log(n)}}
}
{%mechane
\left(\alpha n^{3/4}\right)^{\alpha n^{1/4}}\left(\frac{\alpha n^{3/4}}{log(n)}-\alpha n^{1/4}\right)^{-\alpha n^{1/4}+\frac{\alpha n^{3/4}}{log(n)}}
}=\\

\left(\frac{1-n^{-1/2}}{\frac{1}{\log(n)} -n^{-1/2}}\right)^{\frac{\alpha n^{3/4}}{\log(n)}-\alpha n^{1/4}}\left(\log(n)\right)^{\frac{-\alpha n^{3/4}}{\log(n)}}=\\

\left(\frac{\left(1-n^{-1/2}\right)\log(n)}{1-n^{-1/2}\log(n)}\right)^{\frac{\alpha n^{3/4}}{\log(n)}-\alpha n^{1/4}}\left(\log(n)\right)^{\frac{-\alpha n^{3/4}}{\log(n)}}=\\
A(n)\left(B(n)\right)^{\alpha n^{1/4} -\alpha n^{-1/4}\log(n)}C(n)^{-\alpha n^{-1/4}+\frac{\alpha n^{1/4}}{\log(n)}}\\
\ena
\eeq
where,
\beq
\bea{l}
A(n)=\left(log(n)\right)^{-\alpha n^{1/4}}=e^{-\alpha n^{1/4} log(log(n))}\\
B(n)=\left(1-\frac{1}{n^{1/2}(log(n))^{-1}}\right)^{-n^{1/2}(log(n))^{-1}}\\
C(n)=\left(1-n^{-1/2}\right)^{n^{1/2}}\leq e^{-1}
\ena
\eeq
$B(n)$ monotonically decreases to $e$ and for all $n\geq2$
\beq
B(n)\leq e^{3/2}.
\eeq
Therefore,
\beq
\bea{l}
p(y>\alpha n^{1/4})\leq q\leq e^{-\alpha n^{1/4}(\log\log(n)-\frac{3}{2}+\frac{1}{log(n)}-\frac{1}{\sqrt{n}})}
\ena
\eeq
%The exponent is negative for all $n\geq90$ and therefore $p(y>\alpha n^{1/4})$ is subexponentially decaying to zero.
%%%%%%%%%%%%%%%%%%%%%%%%%%%%%%%%%%%%%%%%%%%%%%%%
%%%%%%%%%%%%%%%%%%%%%%%%%%%%%%%%%%%%%%%%%%%%%%%%
%%%%%%%%%%%%%%%%%%%%%%%%%%%%%%%%%%%%%%%%%%%%%%%%
\section{Chernoff bound on the probability of error of the inner code}
\label{subsec:Chernoff bound of the inner code}
In \ref{sec:decoding with probability} we assumed the worst case that if a row decoded with errors, the entire row is wrong. Therefore, for bounding the probability of block error, we need to bound the probability that more than $\alpha n^{1/4}$ rows are decoded with errors. Denote the number of rows with errors by $\tilde{Z}$. We uniformly bound the probability that a given row will be decoded with errors by $q(n)$ (see (\ref{ineq:Chernoff bound for row error}) ). Define a random  i.i.d binary process $y_i$ with $P(y_i=1)=q$ , $i=1,2,...,n^{1/2}$, set $Z=\displaystyle\sum_{i=1}^{n^{1/2}}{y_i}$ the binomial random variable with a probability of $p=q(n)$. Therefore, $P(\tilde{Z}>\alpha n^{1/4})\leq P(Z>\alpha n^{1/4})$  which can be bound by Chernoff
\beq
\label{ineq:chernoff bound on the probability of error on block appendix}
P(Z>\alpha n^{1/4})\leq e^{-s\alpha n^{1/4}}\left(q e^{s}+1-q\right)^{n^{1/2}}
\eeq
By the first derivative test of the RHS of (\ref{ineq:chernoff bound on the probability of error on block appendix}), one can find that
\beq
s=\log\left(\frac{(1-q)\alpha n^{1/4}}{q(n^{1/2}-\alpha n^{1/4})} \right)
\eeq
and it easy to show that since $\alpha\leq n^{1/4}$, there exists a number $N_0$ (typically a small number) such that for all $n\geq N_0$ we have $s > 0$.
Therefore,
\beq
\bea{l}
\mbox{P(block error)}\leq\frac{(1-q)^{n^{1/2}-\alpha n^{1/4}}(\alpha n^{1/4})^{-\alpha n^{1/4}}(n^{1/2})^{n^{1/2}}}{q^{-\alpha n^{1/4}}(n^{1/2}-\alpha n^{1/4})^{n^{1/2}-\alpha n^{1/4}}}\leq\\
\leq\frac{(\alpha n^{1/4})^{-\alpha n^{1/4}}(n^{1/2})^{n^{1/2}}}{q^{-\alpha n^{1/4}}(n^{1/2}-\alpha n^{1/4})^{n^{1/2}-\alpha n^{1/4}}}\leq\\
q^{\alpha n^{1/4}}\left(\frac{n^{1/4}}{\alpha}-1\right)^{\alpha n^{1/4}}\left(1-\alpha n^{-1/4}\right)^{-n^{1/2}}\leq\\
e^{-\alpha^2 n^{1/2}(\log\log(n)-\frac{3}{2}+\frac{\log(1-\alpha n^{-1/4})}{\alpha^2}-\frac{n^{-1/4}\log(\frac{n^{1/4}}{\alpha}-1)}{\alpha})}.
\ena
\eeq
Note that the exponent is negative for all $n\geq N_0(\alpha)$ (for example $N_0=3340$ for $\alpha=0.65$); therefore the bound decays sub-exponentially in block size $n$.

\bibliographystyle{IEEEtran}
%\bibliography{Turbo_Avi_2009_bib}

\end{document}